\documentclass[preprint,nofootinbib,amsmath,amssymb,aps,prl]{revtex4-1}
\usepackage{graphicx}
\usepackage{float}
\usepackage[T1]{fontenc}
\usepackage[latin1]{inputenc}
\usepackage{amssymb}
\usepackage{subfigure} 
\usepackage{subfig}
\usepackage{float}
\include{epsf}

\newcommand{\bea}{\begin{eqnarray}}
\newcommand{\eea}{\end{eqnarray}}

\makeatother
\begin{document}
\setlength{\unitlength}{1mm}

\title{\Large{Towards Next-to-Leading Order Transport Coefficients from the Four-Particle Irreducible Effective Action}}

\author{M.E. Carrington}
\email{carrington@brandonu.ca}
\affiliation{Department of Physics, Brandon University, Brandon, Manitoba, R7A 6A9 Canada\\ and \\  Winnipeg Institute for Theoretical Physics, Winnipeg, Manitoba }
\author{E. Kovalchuk}
\email{kavalchuke@brandonu.ca}
\affiliation{Department of Physics, Brandon University, Brandon, Manitoba, R7A 6A9 Canada\\ and \\  Winnipeg Institute for Theoretical Physics, Winnipeg, Manitoba }

\begin{abstract}

Transport coefficients can be obtained from 2-point correlators using the Kubo formulae. It has been shown that the full leading order result for electrical conductivity and (QCD) shear viscosity is contained in the re-summed 2-point function that is obtained from the 3-loop 3PI re-summed effective action. The theory produces all leading order contributions without the necessity for power counting, and in this sense it provides a natural framework for the calculation. In this article we study the 4-loop 4PI effective action for a scalar theory with cubic and quartic interactions, with a non-vanishing field expectation value. 
We obtain a set of integral equations that determine the re-summed 2-point vertex function. A next-to-leading order contribution to the viscosity could be obtained from this set of coupled equations. 

\end{abstract}

\maketitle
\section{Introduction}

To date, little progress has been made on the calculation of transport coefficients beyond leading order. A next-to-leading order contribution to shear viscosity in pure $\phi^4$ theory has been calculated in \cite{moorePhi4}. The authors take into account the shift in the soft scalar propagator pole due to the thermal mass. This produces a correction of order $\sqrt{\lambda}$ relative to the leading order result. 
The momentum diffusion coefficient for a non-relativistic heavy quark has been calculated at next-to-leading order in \cite{simon1,simon2}. Hydrodynamic coefficients have been calculated from a kinetic theory approach using a second order gradient expansion, and working at leading order in the coupling \cite{guyYork}. In this paper we propose a new strategy to calculate transport coefficients beyond leading order using an $n$-particle irreducible ($n$PI) effective theory.

An $n$PI  effective theory is defined in terms of $n$ functional arguments which correspond to a set of $n$-point functions that are determined self-consistently through a variational procedure. This procedure re-sums certain classes of diagrams, and represents a re-organisation of perturbation theory (for recent reviews see \cite{bergesReview,bergesSEWM2004}). A well known example of a case in which selective re-summations play an important role in quantum field theory is the hard thermal loop theory, which includes screening effects that regulate infra-red divergences. $n$PI approximation schemes are of particular interest because they can be used to study far from equilibrium systems \cite{cox,aartsNonEq0,smitNonEq,bergesNonEq,aartsNonEq1,aartsNonEq2}. To date, numerical calculations have only been done for 2PI theories where it has been shown that the convergence of perturbative approximations is improved  (see \cite{bergesReview,bergesSEWM2004,bergesConvg} and references therein). 

Recently, it has been demonstrated that $n$PI effective theories provide a natural framework to organise the calculation of transport coefficients. A large $N_f$ leading log  calculation of QED conductivity and shear viscosity was done in \cite{gertNf} using  2PI effective theory. The electrical conductivity and QCD shear viscosity have been obtained at leading order from the 3-loop 3PI effective action \cite{MC-EK1,MC-EK2,MC-EK3}. The formalism produces integral equations for the vertex functions that re-sum the pinching and collinear singularities and produce the full leading order result. In this sense, the 3-loop 3PI effective action respects the natural organisation of the calculation: it produces the full leading order result  without the need for any kind of power counting arguments. Since power counting is notoriously difficult, and becomes increasingly complicated at higher orders, $n$PI effective theories could provide a useful method to organise the calculation of transport coefficients beyond leading order. 

 In spite of the complexity of the $n$PI effective action, a systematic expansion can be done in a self-consistent way. A self-consistently complete loop expansion of the effective action can be based on an  equivalence hierarchy \cite{berges1}: to 2-loop order, the infinite-PI effective action is equivalent to the 2PI effective action, to 3-loop order, the infinite-PI effective action is equivalent to the 3PI effective action, etc. We work with the 4-loop 4PI effective action for a scalar theory with cubic and quartic interactions, with non-vanishing field expectation value. The goal  is to take a first step toward the calculation of next-to-leading order transport coefficients. We derive the integral equations that would produce the next-to-leading order contribution to the viscosity. We remark that the numerical solution of these equations is more difficult than anything that has been accomplished so far. 
 
In addition, there are unresolved issues that would arise if we attempted to extend the calculation to gauge theories \cite{calzettaReview,julienReview}. The renormalizability of a theory is related to the existence of symmetry constraints on the $n$-point functions. For $n$PI effective theories, symmetries and renormalizabilty are connected to the fact that proper $n$-point functions can be defined in more than one way. All definitions are completely equivalent for the exact theory, but they are not the same at finite approximation order.
These issues are well understood for scalar theories and QED at the 2PI level.
For scalar theories one can define a 2-point function that satisfies Goldstone's theorem in the broken phase \cite{baier2PI,vanHees3}. For QED one can define $n$-point functions that obey traditional Ward identities \cite{julienSym,calzettaSym}. These symmetry constraints allow one to construct a complete renormalization that preserves the symmetries of the original theory \cite{vanHees1,vanHees2,vanHees3,reinosaRenorm1,reinosaRenorm2}.
For nonabelian theories, the situation is more involved. It has been shown that at any order in the approximation scheme, the gauge dependence of the effective action always appears at higher approximation order \cite{smit,HZ}. However, the gauge symmetries of the $n$-point functions are more complicated than for abelian theories, and renormalizability remains an open question.

The definitions of the various $n$-point functions that we will use are given below. The symmetry properties of these functions, which are described in the corresponding paragraphs, have not been established for nonabelian gauge theories. 

(1) 1PI vertex functions are obtained by functional differentiation of the 1PI effective action. These functions satisfy the standard symmetry constraints, which reflect the symmetries of the Lagrangian. For example: the 2-point function satisfies Goldstone's theorem for scalar theories in the broken phase, and  the Ward identity for QED (transversality in momentum space). 

(2) Variational $n$-point vertex functions are the functional arguments of the $n$PI effective action. 
Extremising with respect to these functions produces integral equations for the vertices called equations of motion. The vertices can be determined self-consistently by solving the equations of motion using a variational procedure. They do not satisfy standard symmetry constraints.

(3) Re-summed $n$-point vertex functions are obtained by taking functional derivatives of the re-summed effective action, which results from substituting the self-consistent solutions for all variational $n$-point vertex functions with $n\ge 2$ into the $n$PI effective action. Re-summed $n$-point vertex functions have a geometrical interpretation in field space, and satisfy standard symmetry constraints.

(4) Mixed $n$-point functions can be defined by differentiating self-consistent vertices, which define the extremum of the $n$PI effective action, with respect to the field expectation value. Integral equations for these vertex functions are obtained by functionally differentiating the equations of motion of the $n$PI effective action with respect to the field expectation value (see, for example, Eq. (\ref{omegaSt})). For QED, it has been shown that these functions also satisfy standard symmetry constraints \cite{julienSym}. \\

The complete set of integral equations which give the re-summed 2-point vertex function is derived in this paper. Using the Kubo formula that relates the shear viscosity to the 2-point correlator, a next-to-leading order contribution to the shear viscosity could be obtained from this set of coupled equations. The re-summed 2-point function is obtained by functionally differentiating the re-summed effective action. The resulting expression contains  mixed vertex functions. These mixed vertex functions satisfy integral equations which contain variational vertex functions. The variational vertex functions are determined by additional integral equations. 

The paper is organised as follows. In Section \ref{section4PI} we define our notation and give our result for the 4-loop 4PI effective action (some details of the calculation are given in Appendix \ref{appendixA}). In Section \ref{sectionEXT} we calculate the re-summed  2-point vertex function. In Section \ref{sectionBS} we calculate the integral equations satisfied by the mixed vertex functions that appear in the re-summed 2-point vertex function. In Section \ref{sectionUV} we calculate the integral equations satisfied by the variational vertex functions that appear in the mixed vertex functions (that appear in the re-summed 2-point vertex function).

\section{the 4PI Effective Action}
\label{section4PI}

4PI effective actions were introduced in Ref. \cite{deDom1,deDom2}. They were first discussed in the context of relativistic field theories in  Ref.  \cite{norton}. The  3-loop 4PI effective action was calculated in Refs. \cite{berges1,deDom1,deDom2,mec}. The fourth Legendre transform has been studied in \cite{kim}. In this article we study a scalar theory with cubic and quartic interactions, with a non-vanishing field expectation value.   We use the method of successive Legendre transformations \cite{deDom1,berges1} and obtain the 4PI effective action to 4-loop order. The result is given in this section, and some details of the calculation are presented in Appendix \ref{appendixA}. 
We use the following notation:  $D^0$, $U^{oo}$ and $V^0$ are the bare propagator and vertices,  $D$ is the variational propagator, $U$ is the  variational 3-point vertex, and $V$ is the  variational 4-point  vertex. 
We use a compactified notation in which a single numerical subscript represents all space-time co-ordinates. For example: the field expectation value is written $\phi_1:=\phi(x)$, 
the variational propagator is written $D_{12}:=D(x_1,x_2)$, the bare 4-point vertex is written $V^0_{1234}:=V^0(x_1,x_3,x_2,x_4)$, etc.  We also use an Einstein convention in which a repeated index implies an integration over space-time variables. 
We will define propagators and vertices with factors of $i$ so that figures look as simple as possible: 
lines, and intersections of lines, correspond directly to propagators and vertices, with no additional factors of plus or minus $i$.
Using this notation we write the classical action:
\bea
\label{scl}
S_{cl}[\phi]=\frac{1}{2}\phi_1\big[i\,(D^{0}_{12})^{-1}\big]\phi_2-\frac{i}{\;3!} U_{123}^{oo}\phi_1\phi_2\phi_3-\frac{i}{\;4!}V_{1234}^0\phi_1\phi_2\phi_3\phi_4\,.
\eea

We define the effective classical vertex $U^0_{ijk}:=U_{ijk}^{oo}+\phi_l V_{ijkl}^0$ and obtain:
\bea
\label{free}
&&(D^0_{12}(\phi))^{-1}=-i \frac{\delta^2 S_{cl}}{\delta \phi_2\delta \phi_1}\,,\nonumber\\
&&
U^0_{123}=i\frac{\delta^3 S_{cl}}{\delta \phi_3\delta \phi_2\delta \phi_1}= -\frac{\delta (D^{0}_{12}(\phi))^{-1}}{\delta \phi_3}\,,\nonumber\\
&&
V^0_{1234}=i\frac{\delta^4 S_{cl}}{\delta \phi_4 \delta \phi_3\delta \phi_2\delta \phi_1} = \frac{\delta U_{123}^0}{\delta \phi_4}= -\frac{\delta^2 (D^{0}_{12}(\phi))^{-1}}{\delta \phi_4\delta \phi_3}\,.
\eea

The 4PI effective action can be written:
\begin{eqnarray}
\label{Gamma4PI}
\Gamma[\phi,D,U,V]
&&=S_{cl}[\phi]+
    \frac{i}{2} {\rm Tr} \,{\rm Ln}D^{-1}_{12} \nonumber\\
    && +
\frac{i}{2} {\rm Tr}\left[(D^0_{12}(\phi))^{-1}\left(D_{21}-D^0_{21}(\phi)\right)\right]+\Gamma^0[\phi,D,U,V]+\Gamma^{int}[D,U,V] \,.
\eea
The propagator $D$ and vertices $U$ and $V$ are to be determined self-consistently from the equations of motion. The terms $\Gamma^0[\phi,D,U,V]$ and $\Gamma^{int}[D,U,V]$ contain all contributions to the effective action which have two or more loops.  The first piece $\Gamma^0[\phi,D,U,V]$ includes all terms that contain  bare vertices. The calculation of $\Gamma^{int}[D,U,V]$ is lengthy, but straightforward. We show some of the steps in Appendix \ref{appendixA}. The result is shown in Fig. \ref{PHI} \footnote{Figures in this paper are drawn using Jaxodraw.}. 
We define $\Phi = i(\Gamma^0+\Gamma^{int})$ which allows us to represent $\Phi$ as a series of diagrams with all factors of $i$ absorbed into the definitions of the propagators and vertices. 
We have given each diagram in the figure a name, so that we can refer to them individually later.
\par\begin{figure}[H]
\begin{center}
\includegraphics[width=15cm]{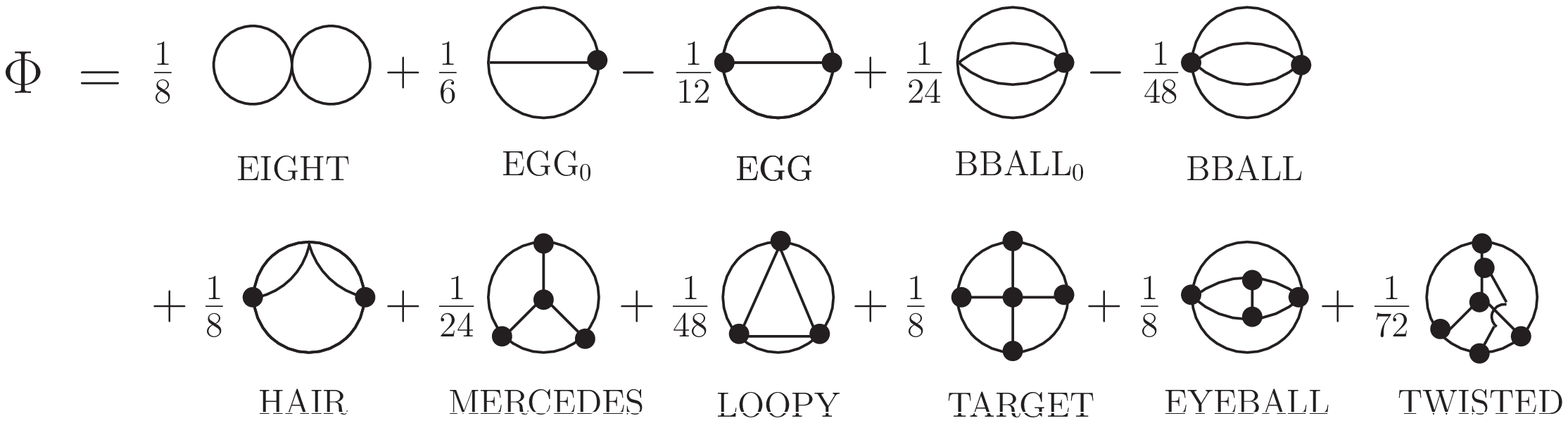}
\end{center}
\caption{\label{PHI}4-loop 4PI effective action.}
\end{figure}
The equations of motion are obtained from the stationarity of the action. There are four equations which are obtained by functionally differentiating with respect to the four functional arguments of the effective action:
\bea
\label{eom1}
&&\frac{\delta \Gamma[\phi,D,U,V]}{\delta X_i}=0\,,~~~~X_i \in \{\phi,D,U,V\}\,.
\eea
The equations obtained by varying with respect to $\{D,U,V\}$ can be solved simultaneously for the self-consistent solutions which are functions of the field expectation value: $\tilde D[\phi]$, $\tilde U[\phi]$, $\tilde V[\phi]$.
Substituting these self-consistent solutions we 
obtain the resummed action, which depends only on the expectation value of the field:
\begin{eqnarray}
\label{Gamma4PI-rs}
\tilde{\Gamma}[\phi]=
\Gamma[\phi, \tilde{D}[\phi], \tilde{U}[\phi], \tilde{V}[\phi]]
\,.
\end{eqnarray}
In the future we will write $\Gamma$ and $\tilde \Gamma$ without their arguments.

\section{$2$-point vertex functions}
\label{sectionEXT}

We start by defining some mixed vertex functions
using the same notation as (\ref{free}):
\begin{eqnarray}
\label{vert-defns}
&&\Omega_{123} = -\frac{\delta \tilde{D}^{-1}_{12}}{\delta \phi_{3}}\,,~~~~\Psi_{1234}=\frac{\delta \tilde U_{123}}{\delta\phi_4}\,.
\end{eqnarray}
These vertices are shown in Fig. \ref{extVerts}. Legs that correspond to functional differentiation with respect to the expectation value of the field are called `external.' These legs are distinguished by an arrow. For both $\Omega$ and $\Psi$ the last index is always assigned to the external leg. In order to simplify the form of the equations we will write $U_0$ as $\Omega_0$ and $V_0$ as $\Psi_0$ throughout the next two sections. 
\par\begin{figure}[H]
\begin{center}
\includegraphics[width=8cm]{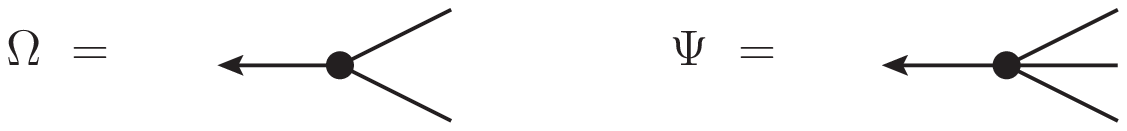}
\end{center}
\caption{\label{extVerts}The vertices $\Omega$ and $\Psi$.}
\end{figure}
\noindent An additional useful relation can be obtained from the identity $\tilde{D}^{-1}_{13}\tilde{D}_{32}=\delta_{12}$. 
Differentiating with respect to $\phi$ and using (\ref{vert-defns}) gives:
\bea
\label{invert}
\frac{\delta\tilde{D}_{12}}{\delta \phi_{3}} = 
\tilde{D}_{11'}\tilde{D}_{22'}  \Omega_{1'2'3}\,.
\eea

The re-summed propagator is defined as: 
\bea
\label{ext-prop}
i(D^{{\rm ext}}_{12})^{-1}=   
     \frac{\delta^2}{\delta \phi_2 \delta \phi_1}
     \tilde{\Gamma}[\phi]\,.
\eea
The re-summed 2-point vertex function, or the re-summed self energy, is extracted from the re-summed propagator using:
\bea
\label{dexternal}
(D^{{\rm ext}}_{12})^{-1}=(D^{0}_{12}(\phi))^{-1}- \Pi^{\rm ext}_{12}\,.
\eea
We can derive an expression for the re-summed 2-point function as a function of the vertices in (\ref{vert-defns}) by taking derivatives of the modified effective action and using the chain rule. 
We use the notation $X_i$ to indicate one of the set of functional variables $X:=\{D,U,V\}$ and $\tilde X_i$ to indicate one of the set of self-consistent solutions $\tilde X:=\{\tilde D[\phi],\,\tilde U[\phi],\,\tilde V[\phi]\}$. We obtain:
\begin{eqnarray}
\label{Dext-long}
i(D^{{\rm ext}}_{12})^{-1} &&
= \frac{\delta^2\Gamma}{\delta \phi_{2} \delta \phi_{1}}\Big|_{\tilde X}+\sum_i \frac{\delta \Gamma}{\delta X_i}\Big|_{\tilde X}\frac{\delta^2 \tilde X_i}{\delta \phi_1\delta \phi_2} \\
&& +\Big[\sum_{i}\frac{\delta^2\Gamma}{\delta X_i\delta \phi_1}\Big|_{\tilde X}\frac{\delta \tilde X_i}{\delta \phi_2}~+~\{1\leftrightarrow 2\}\Big]
 +\sum_{i}\sum_{j}\frac{\delta^2\Gamma}{\delta X_i\delta X_j}\Big|_{\tilde X}\frac{\delta \tilde X_i}{\delta \phi_1}\frac{\delta \tilde X_j}{\delta \phi_2}\,.\nonumber
\end{eqnarray}
The last term in the first line is identically zero (see Eq. (\ref{eom1})). The expression can be further simplified by using the set of equations obtained by differentiating the equations of motion:
\bea
\frac{\delta}{\delta \phi_2}\;\Big[\frac{\delta \Gamma}{\delta X_i}\Big|_{\tilde X}\Big]=0~~\Rightarrow~~
\frac{\delta^2 \Gamma}{\delta X_i\delta \phi_2}\Big|_{\tilde X}+\sum_{j}\frac{\delta^2\Gamma}{\delta X_j\delta X_i}\Big|_{\tilde X}\frac{\delta \tilde X_j}{\delta \phi_2} = 0\,.
\eea
Using this constraint (\ref{Dext-long}) becomes:
\bea
i(D^{{\rm ext}}_{12})^{-1} =\frac{\delta^2\Gamma}{\delta \phi_{2} \delta \phi_{1}}\Big|_{\tilde X}+\sum_{i}\frac{\delta^2\Gamma}{\delta X_i\delta \phi_1}\Big|_{\tilde X}\frac{\delta \tilde X_i}{\delta \phi_2}\,.
\eea
Expanding the sum we have:
\bea
\label{Dext-1}
&& i(D^{{\rm ext}}_{12})^{-1} =\frac{\delta^2\Gamma}{\delta \phi_{2} \delta \phi_{1}}\Big|_{\tilde X}+\frac{\delta^2\Gamma}{\delta D_{34}\delta \phi_1}\Big|_{\tilde X} \frac{\delta \tilde D_{34}}{\delta \phi_2} 
+\frac{\delta^2\Gamma}{\delta U_{345}\delta \phi_1}\Big|_{\tilde X}\frac{\delta \tilde U_{345}}{\delta \phi_2}\,.
\eea
The last term in the sum does not contribute because the derivative $\delta^2 \Gamma/(\delta V\delta\phi)$ is identically zero. Using (\ref{free}), (\ref{Gamma4PI}),  (\ref{vert-defns}) and (\ref{invert}) we obtain:
\bea
\label{Dext-2}
-i \frac{\delta^2\Gamma}{\delta \phi_2 \delta \phi_1} 
&& = (D^0_{12}(\phi))^{-1}-\frac{1}{2}\Psi^0_{1234}\,D_{43}\,, \nonumber\\
-i \frac{\delta^2\Gamma}{\delta D_{43}\delta \phi_1}\Big|_{\tilde X} \frac{\delta \tilde D_{43}}{\delta \phi_2} && = -\frac{1}{2}(\Omega_0^\prime)_{341}\cdot \big(\tilde D_{33^\prime}\tilde D_{44^\prime}\Omega_{3^\prime 4^\prime 2}\big)\,,~~\Omega_0^\prime=\Omega_0+2\frac{\delta^2 \Phi}{\delta D\,\delta \phi}\,,\nonumber\\
-i \frac{\delta^2\Gamma}{\delta U_{543}\delta \phi_1}\Big|_{\tilde X} \frac{\delta \tilde U_{543}}{\delta \phi_2} && = -\frac{1}{6}\Psi^0_{3^\prime4^\prime5^\prime1}\cdot\big(\tilde D_{3^\prime3}D_{4^\prime4}D_{5^\prime5}\Psi_{3 4 5 2}\big)\,.
\eea
In the first line of Eq. (\ref{Dext-2}), the two contributions on the right  side come from the classical and 1-loop part of $\Gamma$ (the first three terms on the right  side of (\ref{Gamma4PI})). In the second line we have used $\delta/\delta D_{43}(\delta  \Phi/\delta \phi_1)=\delta/\delta D_{43}(\delta  \Phi_{{\rm egg}_0}/\delta \phi_1)$ and defined the vertex $\Omega^\prime_0$: 
\bea
-i \frac{\delta^2\Gamma}{\delta D_{43}\delta \phi_1}=-\frac{1}{2}\Omega^0_{341}-\frac{\delta^2  \Phi_{{\rm egg}_0}}{\delta D_{43}\delta \phi_1}=: -\frac{1}{2}(\Omega^\prime_0)_{341}\,,
\eea
where the term $\Omega^0_{341}$ comes from the 1-loop part of $\Gamma$. The vertex $\Omega_0^\prime$ is shown in the first two graphs on the right side of the first line of Fig. \ref{OMPSint}. In the third line of Eq. (\ref{Dext-2}) we have used:
\bea
-i \frac{\delta^2\Gamma}{\delta U_{543}\delta \phi_1}=- \frac{\delta^2\Phi_{\rm egg_0}}{\delta U_{543}\delta \phi_1} = -\frac{1}{6}\Psi^0_{3^\prime 4^\prime5^\prime1^\prime}D_{3^\prime3}D_{4^\prime4}D_{5^\prime5}\,.
\eea
If the future we will use the shorthand notation:
\bea
&& \tilde D_{33^\prime}\tilde D_{44^\prime}\Omega_{3^\prime  4^\prime2} = (\tilde D\,\tilde D\,\Omega)_{342}\,,\nonumber\\
&& \tilde D_{33^\prime}D_{44^\prime}D_{55^\prime}\Psi_{3^\prime 4^\prime 5^\prime 2} = (\tilde D\tilde D\tilde D\Psi)_{3452}\,.
\eea
We extract $\Pi_{12}^{\rm ext}$ from (\ref{dexternal}), (\ref{Dext-1}) and (\ref{Dext-2}):
\bea
\label{Dext-3}
\Pi_{12}^{\rm ext} &&= \frac{1}{2}\Psi^0_{1234}\,D_{43}
+\frac{1}{2}\big(\Omega_0^\prime\big)_{341} (\tilde D\,\tilde D\,\Omega)_{432}
+ \frac{1}{6} \Psi^0_{3451} (\tilde D\tilde D\tilde D\Psi)_{5432}\,.
\eea
The result agrees with the Schwinger-Dyson equation for the 2-point vertex function  \cite{cvitanovic,kajantie} and is shown in Fig. \ref{PIextdiag}.
\par\begin{figure}[H]
\begin{center}
\includegraphics[width=13cm]{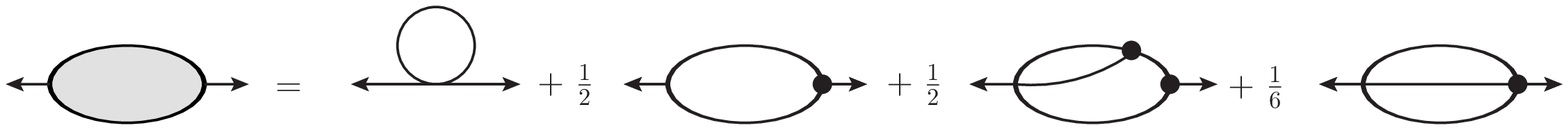}
\end{center}
\caption{\label{PIextdiag}The re-summed 2-point vertex function.}
\end{figure}

The variational 2-point function satisfies an integral equation obtained from the  equation of motion (\ref{eom1}):
\bea
2\frac{\delta \Gamma}{\delta D_{21}} = 0~~~~\rightarrow~~~~ (D_{12})^{-1}=(D^{0}_{12}(\phi))^{-1}- \Pi_{12}\,.
\eea
By rearranging terms one can show that the variational 2-point vertex function satisfies the same Schwinger-Dyson integral equation as the resummed 2-point vertex function (Eq. (\ref{Dext-3}) and Fig. \ref{PIextdiag}), with the vertices $\Omega$ and $\Psi$ replaced by $U$ and $V$ respectively.

\section{Integral Equations for Mixed Vertex Functions}
\label{sectionBS}

In this section we derive the integral equations for the mixed vertex functions $\Omega$ and $\Psi$ which appear in the re-summed 2-point vertex function $\Pi^{\rm ext}$ (see Fig. \ref{PIextdiag}). In addition to the vertices defined in (\ref{vert-defns}), we will need to define the 5-point vertex: 
\bea
\label{theta-defn}
\Theta_{12345}=\frac{\delta \tilde V_{1234}}{\delta\phi_5}\,.
\eea
As in Eq. (\ref{vert-defns}), the last index is always assigned to the external leg. This vertex is shown in the left side of the last line of Fig. \ref{OMPSint}. The integral equations satisfied by the vertices $\Omega$, $\Psi$  and $\Theta$ are obtained by taking functional derivatives with respect to the field expectation value of the appropriate equations of motion (\ref{eom1}):
\bea
\label{omegaSt}
&&2\frac{\delta}{\delta \phi_9}\bigg[\frac{\delta \Gamma}{\delta D_{12}}\bigg|_{\tilde X}\bigg]=0\,,\\
\label{psiSt}
&&3!\frac{\delta}{\delta \phi_9}\bigg[\frac{\delta \Gamma}{\delta U_{1^\prime2^\prime3^\prime}}\bigg|_{\tilde X}\tilde D^{-1}_{1^\prime 1}\tilde D^{-1}_{2^\prime 2}\tilde D^{-1}_{3^\prime 3}\bigg]=0\,,\\
\label{thetaSt}
&&4!\frac{\delta}{\delta \phi_9}\bigg[\frac{\delta \Gamma}{\delta V_{1^\prime2^\prime3^\prime4^\prime}}\bigg|_{\tilde X}\tilde D^{-1}_{1^\prime 1}\tilde D^{-1}_{2^\prime 2}\tilde D^{-1}_{3^\prime 3} \tilde D^{-1}_{4^\prime 4}\bigg]=0\,.
\eea
The subscript $\tilde X$ indicates that all self-consistent solutions are substituted. The numerical factors 2!, 3! and 4! are inserted for later convenience, as explained below. The index `9' on the $\phi$ derivative is chosen so that this index remains the last (highest) after using the chain rule (see Eqs. (\ref{betaS-A}), (\ref{betaS-B}), (\ref{betaS-C})). In Eq. (\ref{psiSt}) we multiply by three inverse propagators to remove the external legs that appear when we take the derivative with respect to $U$, and in Eq. (\ref{thetaSt}) we multiply by four inverse propagators to remove the external legs that appear when we take the derivative with respect to $V$. 

The next step is to expand the derivatives in (\ref{omegaSt}), (\ref{psiSt}) and (\ref{thetaSt}) using the chain rule. 
We start with (\ref{omegaSt}). For the 1-loop part of the effective action, we take the derivatives explicitly. Using (\ref{vert-defns}) we obtain (including the factor of 2 that is introduced in (\ref{omegaSt}) for this purpose):  $-\Omega_{129}+(\Omega^\prime_0)_{129}$. Rearranging we have:
\bea
\label{betaS-A}
\Omega_{129} &&= (\Omega^\prime_0)_{129} +2\frac{\delta \tilde D_{34}}{\delta \phi_9}\,\bigg[\frac{\delta^2\,\Phi}{\delta D_{34}\delta D_{12}}\bigg|_{\tilde X}\bigg] \nonumber\\
&&
+2\frac{\delta \tilde U_{345}}{\delta \phi_9}\,\bigg[\frac{\delta^2\,\Phi}{\delta U_{345}\,\delta D_{12}}\bigg|_{\tilde X}\bigg]
+2\frac{\delta \tilde V_{3456}}{\delta \phi_9}\,\bigg[\frac{\delta^2\,\Phi}{\delta V_{3456}\,\delta D_{12}}\bigg|_{\tilde X}\bigg]\,.
\eea
Now we consider (\ref{psiSt}). The 1-loop part of the effective action gives zero contribution, since it does not depend on the vertex $U$. Using (\ref{vert-defns}) the result can be written:
\bea
\label{betaS-B}
\Psi_{1239} && = \Psi^0_{1239}
+3! \frac{\delta \tilde D_{45}}{\delta \phi_9}\,\frac{\delta}{\delta \tilde D_{45}}\bigg[\frac{\delta\,\Phi}{\delta U_{1^\prime2^\prime 3^\prime}}\bigg|_{\tilde X}\tilde D^{-1}_{1^\prime 1}\tilde D^{-1}_{2^\prime 2}\tilde D^{-1}_{3^\prime 3}\bigg]\nonumber\\
&&+3! \frac{\delta \tilde U_{456}}{\delta \phi_9}\,\frac{\delta}{\delta \tilde U_{456}}\,\bigg[\frac{\delta\,\Phi^\prime}{\delta U_{1^\prime2^\prime 3^\prime}}\bigg|_{\tilde X}\tilde D^{-1}_{1^\prime 1}\tilde D^{-1}_{2^\prime 2}\tilde D^{-1}_{3^\prime 3}\bigg]\nonumber\\
&&+3! \frac{\delta \tilde V_{4567}}{\delta \phi_9}\,\frac{\delta}{\delta \tilde V_{4567}}\,\bigg[\frac{\delta\,\Phi}{\delta U_{1^\prime2^\prime 3^\prime}}\bigg|_{\tilde X}\tilde D^{-1}_{1^\prime 1}\tilde D^{-1}_{2^\prime 2}\tilde D^{-1}_{3^\prime 3} \bigg]
\,.
\eea
The notation $\Phi^\prime$ indicates that the diagram labelled EGG in Fig. \ref{PHI} has been removed. This  diagram, when multiplied by the factor 3!, produces the term $-\Psi_{1239}$. 
In addition, we have used:
\bea
3!\frac{\delta}{\delta\phi_9}\bigg[\frac{\delta^2\,\Phi}{\delta U_{1^\prime2^\prime 3^\prime}}\bigg|_{\tilde X}\tilde D^{-1}_{1^\prime 1}\tilde D^{-1}_{2^\prime 2}\tilde D^{-1}_{3^\prime 3}\bigg]=3!\frac{\delta}{\delta\phi_9}\bigg[\frac{\delta^2\,\Phi_{\rm EGG_0}}{\delta U_{1^\prime2^\prime 3^\prime}}\bigg|_{\tilde X}\tilde D^{-1}_{1^\prime 1}\tilde D^{-1}_{2^\prime 2}\tilde D^{-1}_{3^\prime 3}\bigg]=\Psi^0_{1239}\,.
\eea
Now we look at (\ref{thetaSt}). The 1-loop part of the effective action gives zero contribution, since it does not depend on the vertex $V$. Using (\ref{vert-defns}) the result can be written:
\bea
\label{betaS-C}
\Theta_{12349}  &&= 
4! \frac{\delta \tilde D_{56}}{\delta \phi_9}\,\frac{\delta}{\delta \tilde D_{56}}\bigg[\frac{\delta\,\Phi}{\delta V_{1^\prime2^\prime 3^\prime 4^\prime}}\bigg|_{\tilde X}\tilde D^{-1}_{1^\prime 1}\tilde D^{-1}_{2^\prime 2}\tilde D^{-1}_{3^\prime 3} \tilde D^{-1}_{4^\prime 4}\bigg]\nonumber\\
&&+4! \frac{\delta \tilde U_{567}}{\delta \phi_9}\,\frac{\delta}{\delta \tilde U_{567}}\,\bigg[\frac{\delta\,\Phi}{\delta V_{1^\prime2^\prime 3^\prime 4^\prime}}\bigg|_{\tilde X}\tilde D^{-1}_{1^\prime 1}\tilde D^{-1}_{2^\prime 2}\tilde D^{-1}_{3^\prime 3} \tilde D^{-1}_{4^\prime 4}\bigg]\nonumber\\
&& +4! \frac{\delta \tilde V_{5678}}{\delta \phi_9}\,\frac{\delta}{\delta \tilde V_{5678}}\,\bigg[\frac{\delta\,\Phi^{\prime\prime}}{\delta V_{1^\prime2^\prime 3^\prime 4^\prime}}\bigg|_{\tilde X}\tilde D^{-1}_{1^\prime 1}\tilde D^{-1}_{2^\prime 2}\tilde D^{-1}_{3^\prime 3} \tilde D^{-1}_{4^\prime 4}\bigg]
\,.
\eea
The notation $\Phi^{\prime\prime}$ indicates that the diagram labelled BBALL in Fig. \ref{PHI} has been removed. This  diagram, when multiplied by the factor 4!, produces the term $-\Theta_{12349}$. 

To simplify the resulting expressions, we introduce the following  notation:
\bea
\label{Mdefn}
&& \frac{\delta^2\Phi}{\delta D_{34}\delta D_{12}}= \frac{1}{\;2!}\frac{1}{\;2!}C_{34;12}\,,\nonumber\\
&&\frac{\delta}{\delta U_{345}}\frac{\delta\Phi}{\delta D_{12}}=\frac{1}{\;2!}\frac{1}{\;3!}D_{33^\prime}D_{44^\prime}D_{55^\prime}C_{3^\prime4^\prime5^\prime;12}\,,\nonumber\\
&&\frac{\delta}{\delta V_{3456}}\frac{\delta\Phi}{\delta D_{12}}=\frac{1}{\;2!}\frac{1}{\;4!}D_{33^\prime}D_{44^\prime}D_{55^\prime}D_{66^\prime}C_{3^\prime4^\prime5^\prime6^\prime;12}
+
\frac{1}{\;3!}\delta_{13}D_{44^\prime}D_{55^\prime}D_{66^\prime}N_{4^\prime5^\prime6^\prime;2}\,,\nonumber\\[4mm]
&& \frac{\delta}{\delta D_{12}}\left[\frac{\delta \Phi}{\delta U_{3^\prime4^\prime5^\prime}}D^{-1}_{3^\prime 3}D^{-1}_{4^\prime 4}D^{-1}_{5^\prime 5}\right] = \frac{1}{\;2!}\frac{1}{\;3!}C_{12;345}\,,\nonumber\\
&&\frac{\delta}{\delta U_{456}} \left[\frac{\delta\Phi^\prime}{\delta U_{1^\prime2^\prime3^\prime}}  D^{-1}_{1^\prime 1} D^{-1}_{2^\prime 2} D^{-1}_{3^\prime 3} \right] =\frac{1}{\;3!}\frac{1}{\;3!} D_{44^\prime}D_{55^\prime}D_{66^\prime}C_{4^\prime5^\prime6^\prime;123}+ \frac{1}{\;2!}\frac{1}{\;2!}\delta_{14}D_{55^\prime}D_{66^\prime}N_{5^\prime6^\prime;23}\nonumber\,,\\
&& \frac{\delta}{\delta V_{4567}} \left[\frac{\delta\Phi^\prime}{\delta U_{1^\prime2^\prime3^\prime}}  D^{-1}_{1^\prime 1} D^{-1}_{2^\prime 2} D^{-1}_{3^\prime 3} \right] = \frac{1}{\;2!}\frac{1}{\;3!}\delta_{14}D_{5^\prime5}D_{6^\prime6}D_{7^\prime7}N_{5^\prime6^\prime7^\prime;23}\nonumber\,,\\[4mm]
&&\frac{\delta}{D_{56}}\left[\frac{\delta\Phi}{V_{1^\prime2^\prime3^\prime4^\prime}}D_{1^\prime1}^{-1}D_{2^\prime2}^{-1}D_{3^\prime3}^{-1}D_{4^\prime4}^{-1}\right]
 = \frac{1}{\;2!}\frac{1}{\;4!}C_{56;1234}\,,\nonumber\\
&& \frac{\delta}{U_{567}}\left[\frac{\delta\Phi}{V_{1^\prime2^\prime3^\prime4^\prime}}D_{1^\prime1}^{-1}D_{2^\prime2}^{-1}D_{3^\prime3}^{-1}D_{4^\prime4}^{-1}\right]
 = \frac{1}{\;2!}\frac{1}{\;3!}\delta_{13}D_{66^\prime}D_{77^\prime}N_{6^\prime7^\prime;234}\,,\nonumber\\
&&\frac{\delta}{V_{5678}}\left[\frac{\delta\Phi}{V_{1^\prime2^\prime3^\prime4^\prime}}D_{1^\prime1}^{-1}D_{2^\prime2}^{-1}D_{3^\prime3}^{-1}D_{4^\prime4}^{-1}\right]
 = \frac{1}{\;2!}\frac{1}{\;2!}\frac{1}{2}\delta_{15}\delta_{26}D_{77^\prime}D_{88^\prime}N_{7^\prime8^\prime;34}\,.
 \eea
For each vertex, the semi-colon divides legs that attach to the left and right sides of the diagram. 
Using this notation and Eqs. (\ref{vert-defns}) and (\ref{theta-defn}), the equations in (\ref{betaS-A}),  (\ref{betaS-B}) and (\ref{betaS-C}) can be written:
\bea
\label{sdOmega}
\Omega_{129}&&=(\Omega^\prime_0)_{129}+\frac{1}{2}\Omega_{34 9}D_{33^\prime}D_{44^\prime}C_{3^\prime4^\prime;12}\nonumber\\
&&+\frac{1}{\;3!}\Psi_{3459}D_{33^\prime}D_{44^\prime}D_{55^\prime}  C_{3^\prime4^\prime5^\prime;12}
+\Theta_{34569}\frac{1}{\;4!}D_{33^\prime}D_{44^\prime}D_{55^\prime}D_{66^\prime}C_{3^\prime4^\prime5^\prime6^\prime;12}\,,
\\
\label{sdPsi}
\Psi_{1239}&&=\Psi^0_{1239}+\frac{1}{2}\Omega_{45 9}D_{44^\prime}D_{55^\prime}C_{4^\prime5^\prime;123}\nonumber\\
&&+\Psi_{4569}
\left[\frac{1}{\;3!}D_{44^\prime}D_{55^\prime}D_{66^\prime} C_{4^\prime5^\prime6^\prime;123}+ \frac{3}{\;2!}\delta_{14}D_{55^\prime}D_{66^\prime}N_{5^\prime6^\prime;23}\right]\nonumber\\
&&+\Theta_{45679} \frac{3}{\;3!}\delta_{14}D_{55^\prime}D_{66^\prime}D_{77^\prime}N_{5^\prime6^\prime7^\prime;23}\,,
\\
\Theta_{12349}&&=\frac{1}{2}\Omega_{56 9}D_{55^\prime}D_{66^\prime}C_{5^\prime6^\prime;1234}+\frac{4}{\;2!} \Psi_{5679}
\delta_{15}D_{66^\prime}D_{77^\prime}N_{6^\prime7^\prime;234}\nonumber\\
&&+ \frac{4!}{2!2!}\frac{1}{2} \Theta_{56789}\delta_{15}\delta_{26}D_{77^\prime}D_{88^\prime}N^\prime_{7^\prime8^\prime;34}\,.
\label{sdTheta}
\eea
\noindent Eqs. (\ref{sdOmega}), (\ref{sdPsi}) and (\ref{sdTheta}) are represented diagrammatically in Fig. \ref{OMPSint}. 
\par\begin{figure}[H]
\begin{center}
\includegraphics[width=18cm]{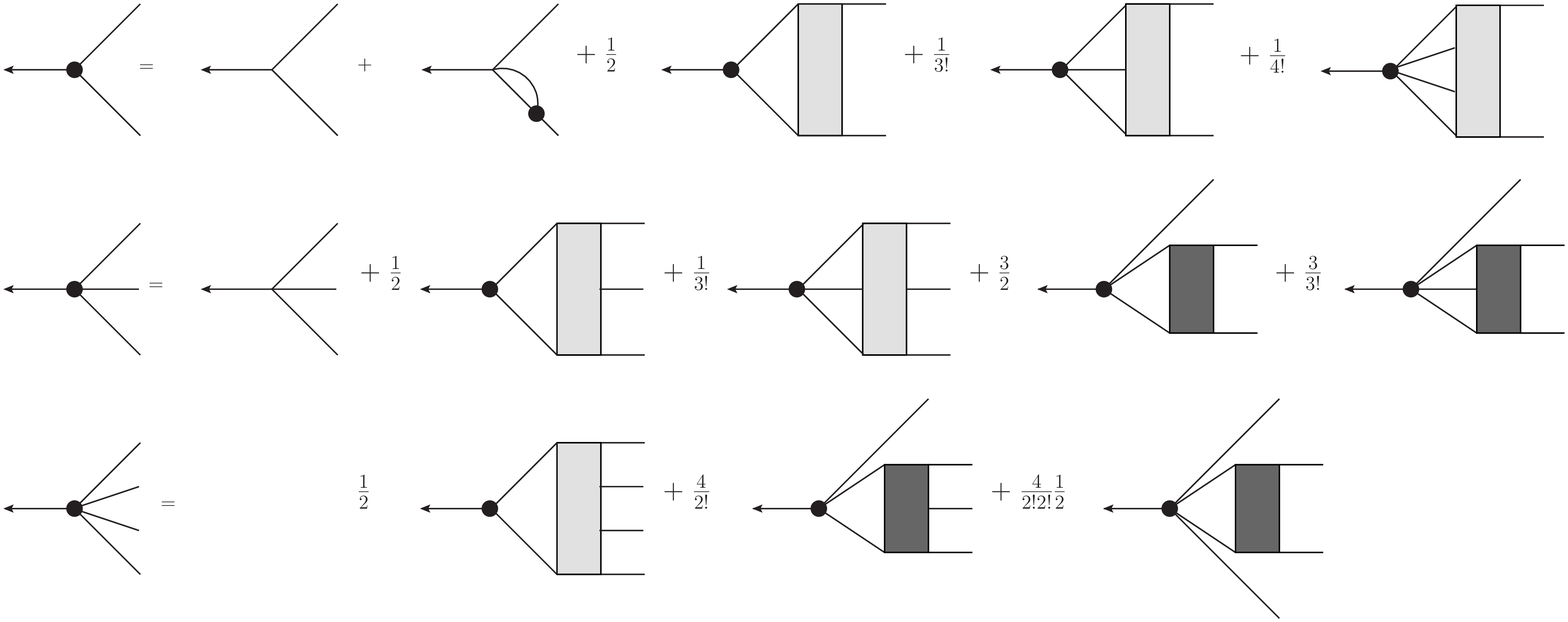}
\end{center}
\caption{\label{OMPSint}The integral equations for the vertices $\Omega$, $\Psi$ and $\Theta$. The grey boxes are the vertices listed in Eq. (\ref{supress}).}
\end{figure}
We note that in Eqs. (\ref{sdOmega}), (\ref{sdPsi}) and (\ref{sdTheta}) and Fig. \ref{OMPSint} we have combined terms that correspond to permutations of external legs. For example, the factor 3 multiplying the fourth term on the right hand side of (\ref{sdPsi}) indicates that three terms have been combined as indicated in Eq. (\ref{extP}):
\bea
\label{extP}
&& \frac{1}{2}\Psi_{4519} D_{44^\prime}D_{55^\prime} N_{4^\prime5^\prime;23}
+\frac{1}{2}\Psi_{4529}D_{44^\prime}D_{55^\prime} N_{4^\prime5^\prime;13}\nonumber\\
&&+\frac{1}{2}\Psi_{4539}D_{44^\prime}D_{55^\prime} N_{4^\prime5^\prime;12}~\Rightarrow~ \frac{3}{2}\Psi_{4539}D_{44^\prime}D_{55^\prime} N_{4^\prime5^\prime;12}\,.\nonumber
\eea

We introduce a shorthand notation for the vertices defined in (\ref{Mdefn}) by suppressing indices. We write:
\bea
\label{supress}
\begin{array}{ccccc}
C_{ij;xy}:=C_{22}\,,& C_{ijk;xy}:=C_{32}\,, & C_{ij;xyz}:=C_{23}\,, & C_{ijk;xyz}:=C_{33}\,, & C_{ijkl;xy}:=C_{42}\,,\\
C_{ij;xyzw}:=C_{24}\,,&
 N_{ij,xy}:=N_{22}\,,& N_{ijk,xy}:=N_{32}\,,& N_{ij,xyz}:=N_{23}\,. &
 \end{array}
\eea
 These vertices are calculated by substituting $\Phi$ (as shown in Fig. \ref{PHI}) into (\ref{Mdefn}). In each case, the symmetric partner of a vertex is obtained by inverting the legs. For example, $C_{23}$ is the right/left inverse of $C_{32}$. To illustrate the procedure we show below the contributions to each of the vertices from the TARGET diagram. \\

We obtain the vertex $C_{22}$ by differentiating twice with respect to $D$. This corresponds to removing two propagators. Different contributions to $C_{22}$ arise depending on which two propagators are removed. The result is shown in Fig. \ref{Mtarget}.
\par\begin{figure}[H]
\begin{center}
\includegraphics[width=11cm]{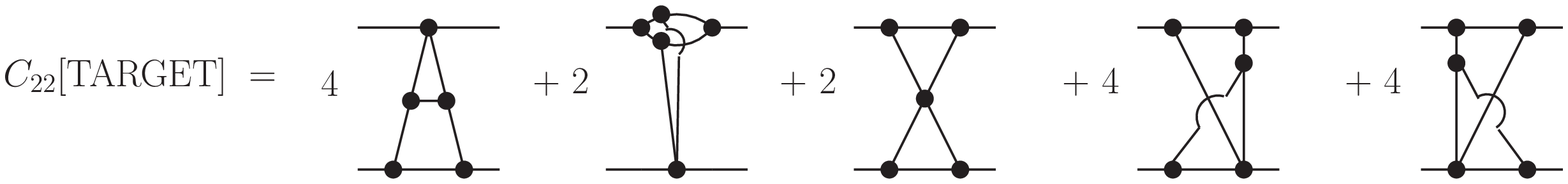}
\end{center}
\caption{\label{Mtarget}Contributions to the vertex $C_{22}$ from the TARGET graph.}
\end{figure}
\noindent The first graph represents 4 different diagrams that have all been combined in the figure to simplify the notation. These 4 diagrams correspond to the 2! 2! = $4$ permutations of the legs on the left and right sides of the figure. The second graph contains a factor 2 instead of 4 because an additional factor of 1/2 is contributed by the symmetry factor of the graph. This symmetry factor comes from the fact that the diagram is symmetric under interchange of the two internal vertices. For the third diagram the factor is 2! 2! $\cdot~\frac{1}{2}$, where the $\frac{1}{2}$ comes because permuting the left legs is equivalent to permuting the right legs (one can see this immediately from the fact that the diagram is symmetric when inverted top-to-bottom). The factor for the fourth and fifth graphs is obtained in the same way as for the first graph. 

Next we consider the vertex $C_{32}$. First, we differentiate with respect to $U$ and remove the attached legs. Then we differentiate with respect to $D$. The results are shown in Fig. \ref{Ctarget}. 
\par\begin{figure}[H]
\begin{center}
\includegraphics[width=10cm]{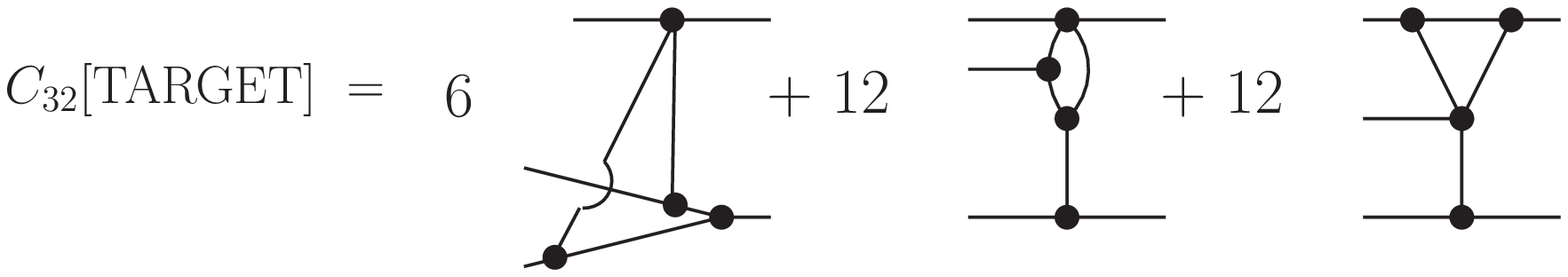}
\end{center}
\caption{\label{Ctarget}Contributions to the vertex $C_{32}$ from the TARGET graph.}
\end{figure}
\noindent The second and third graphs carry a factor 3! 2! = 12 that corresponds to the permutations of legs on the left and right side of the diagram. The first graph has an additional factor of 1/2 from the symmetry factor of the graph, which comes from the fact that the diagram is symmetric under interchange of the two internal vertices.

Next we calculate the contributions to the vertices $C_{33}$ and $N_{22}$. We differentiate with respect to $U$ and remove the attached legs, and then differentiate with respect to $U$ again. In general, these derivatives produce two classes of terms: for some terms there is a delta function which ties together two of the legs (as in the term on the right side of the last line in Eq. (\ref{Mdefn})), and for some terms this delta function does not appear. The results are shown in Fig. \ref{CNtarget}.
\par\begin{figure}[H]
\begin{center}
\includegraphics[width=9cm]{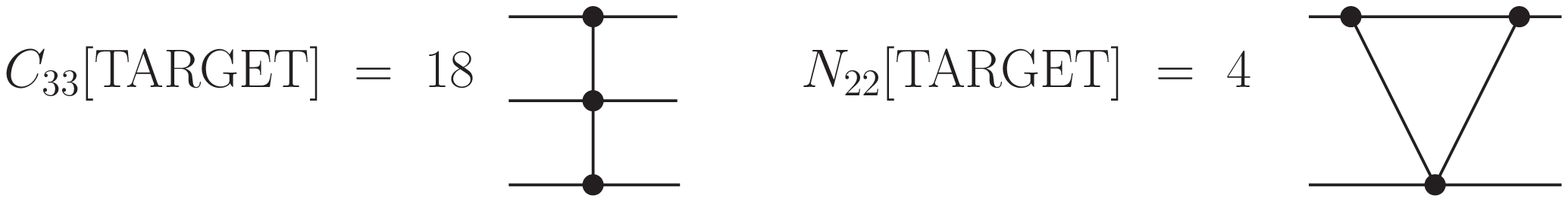}
\end{center}
\caption{\label{CNtarget}Contributions to the vertices $C_{33}$ and $N_{22}$ from the TARGET graph.}
\end{figure}
\noindent The first graph carries a factor 3! 3! $\cdot~\frac{1}{2}=18$ which comes from the possible permutations of legs on the left and right side of the diagram, with the $\frac{1}{2}$ accounting for the fact that the diagram is symmetric when inverted top-to-bottom. The second graph has a factor 2! 2! = 4 from the permutations of the left and right legs. 

Next we calculate the contributions to the vertex $C_{42}$.  We differentiate with respect to $D$, and then differentiate with respect to $V$. The result is shown in Fig. \ref{C42target}.
\par\begin{figure}[H]
\begin{center}
\includegraphics[width=5cm]{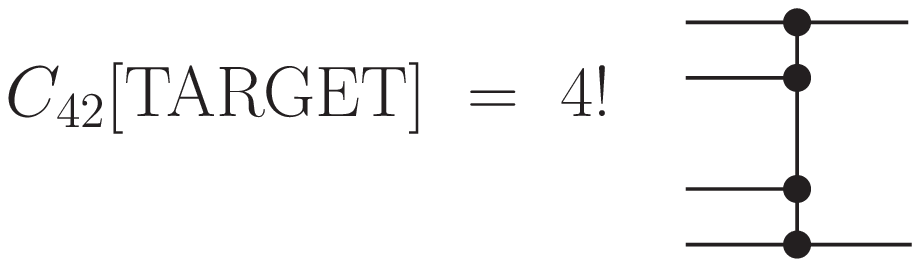}
\end{center}
\caption{\label{C42target}Contribution to the vertex $C_{42}$ from the TARGET graph.}
\end{figure}
Finally we calculate the contributions to the vertex $N_{32}$. We differentiate with respect to $U$ and remove the attached legs, and then differentiate with respect to $V$. The result is shown in Fig. \ref{N32target}.
\par\begin{figure}[H]
\begin{center}
\includegraphics[width=5cm]{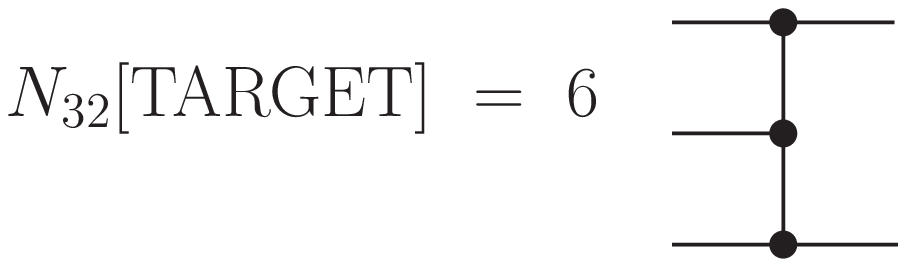}
\end{center}
\caption{\label{N32target}Contribution to the vertex $N_{32}$ from the TARGET graph.}
\end{figure}

The calculation for each diagram in Fig. \ref{PHI} is done in the same way as for the target diagram. Each graph contains a numerical factor that corresponds to the symmetry factor of the diagram, times the permutation factor for the left legs and right legs.  When a vertex is substituted into one of the integral equations (\ref{omegaSt}) or (\ref{psiSt}), the left legs become internal ones. To present the results so that the numerical factors have a more recognisable form, we divide each vertex by the permutation factor for the left legs. Using this notation, the complete results for each of the vertices are shown in Figs. \ref{Meqn} - \ref{N32eqn}. The results have been compactified by using the equation represented in Fig. \ref{Uint} to remove the bare vertex $U_0$. This substitution results in the cancellation of a large number of graphs. 
\par\begin{figure}[H]
\begin{center}
\includegraphics[width=16cm]{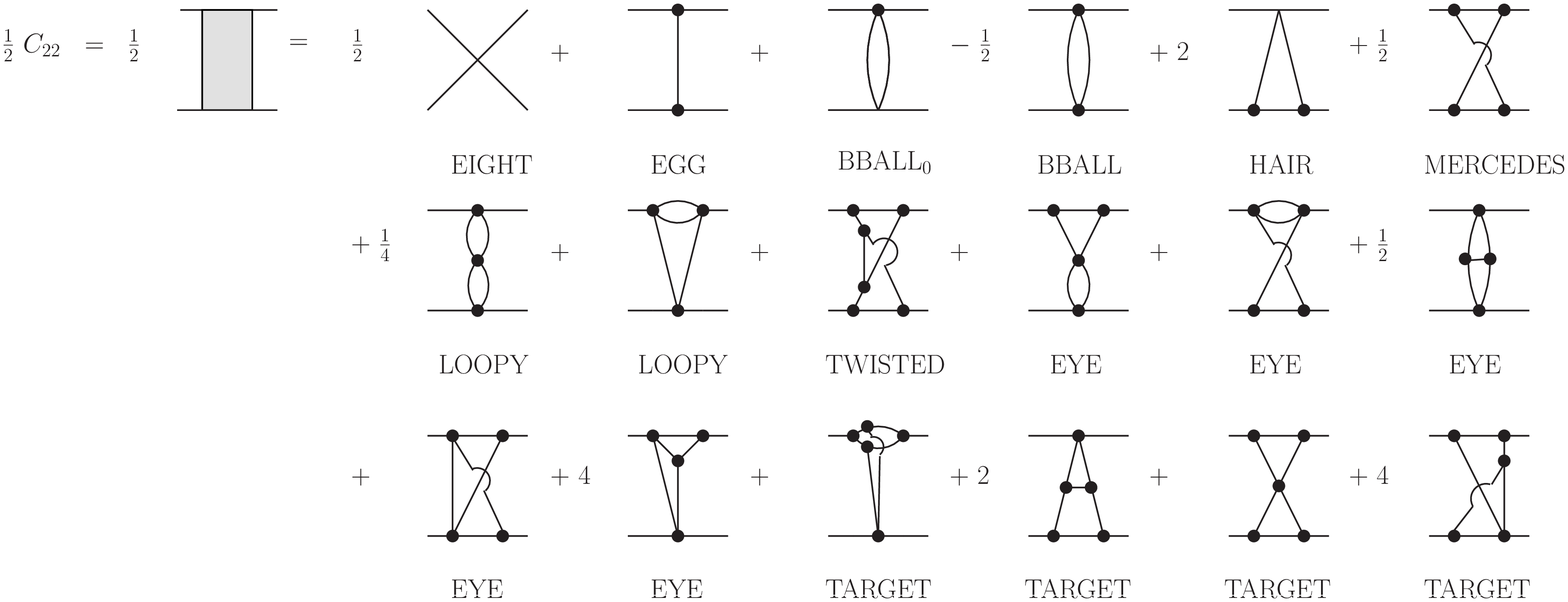}
\end{center}
\caption{\label{Meqn}The shaded box is the vertex $C_{22}$ which is part of the kernel of the $\Omega$ integral equation. The last TARGET graph and the last two EYE diagrams should each be drawn as two graphs where the second one is the left-right inversion of the one in the figure. The graphs are combined to simplify the figure.}
\end{figure}
\par\begin{figure}[H]
\begin{center}
\includegraphics[width=12cm]{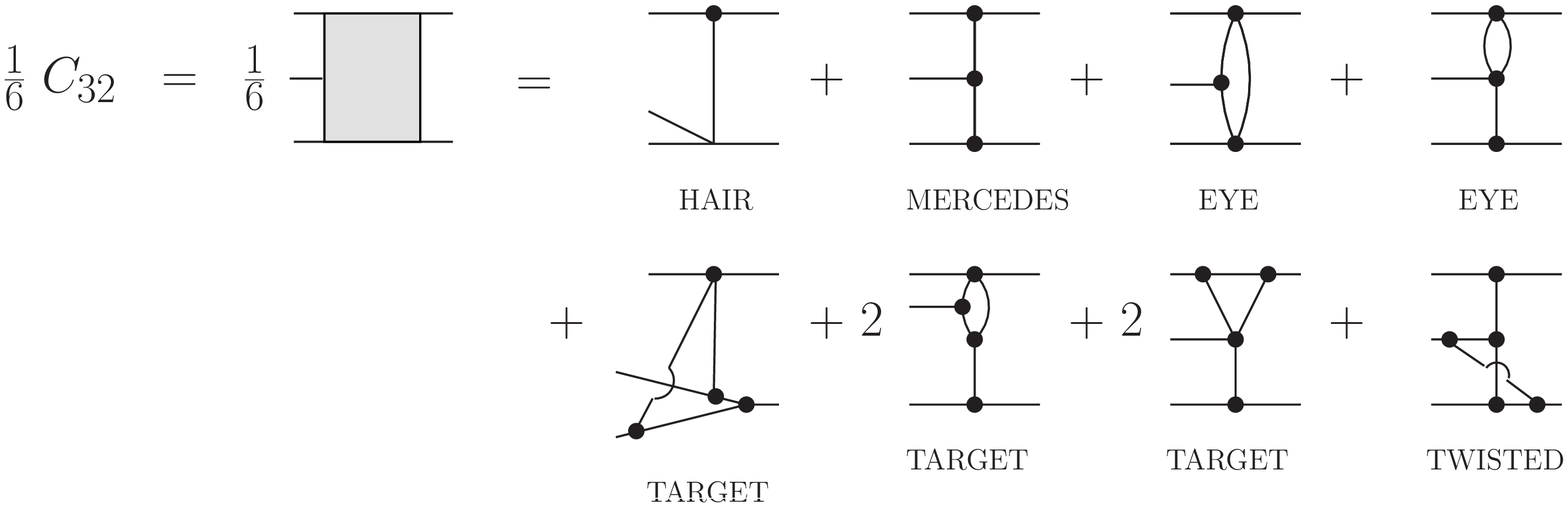}
\end{center}
\caption{\label{Ceqn}The shaded box is the vertex $C_{32}$ which is part of the kernel of the $\Omega$ integral equation. The vertex $C_{23}$ which appears in the $\Psi$ integral equation is obtained by inverting the diagram left to right.}
\end{figure}
\par\begin{figure}[H]
\begin{center}
\includegraphics[width=7cm]{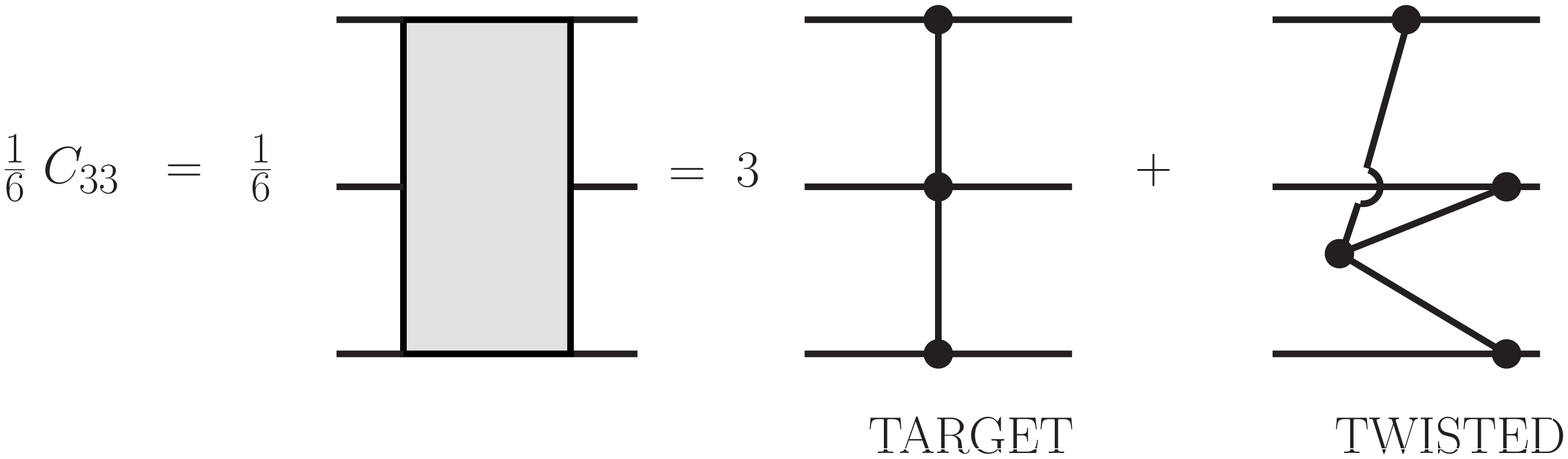}
\end{center}
\caption{\label{Cbareqn}The shaded box is the vertex $C_{33}$ which is part of the kernel of the $\Psi$ integral equation.}
\end{figure}
\par\begin{figure}[H]
\begin{center}
\includegraphics[width=14cm]{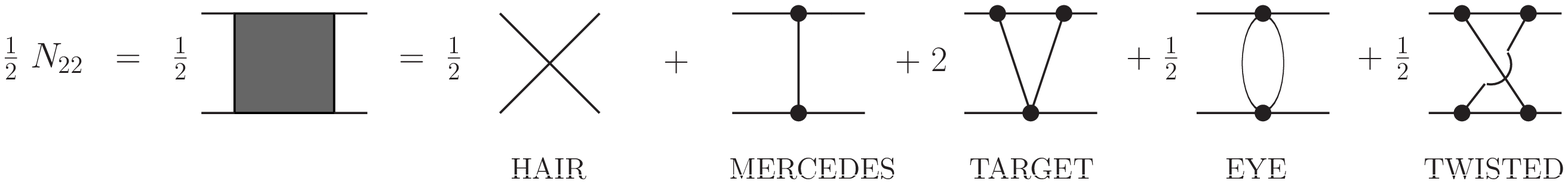}
\end{center}
\caption{\label{Neqn}The shaded box is the vertex $N_{22}$ which is part of the kernel of the $\Psi$ integral equation. The function $N^\prime_{22}$ which appears in the kernel of the $\Theta$ integral equation is given by the first 2 diagrams on the right side, with a full vertex on the first diagram. }
\end{figure}
\par\begin{figure}[H]
\begin{center}
\includegraphics[width=12cm]{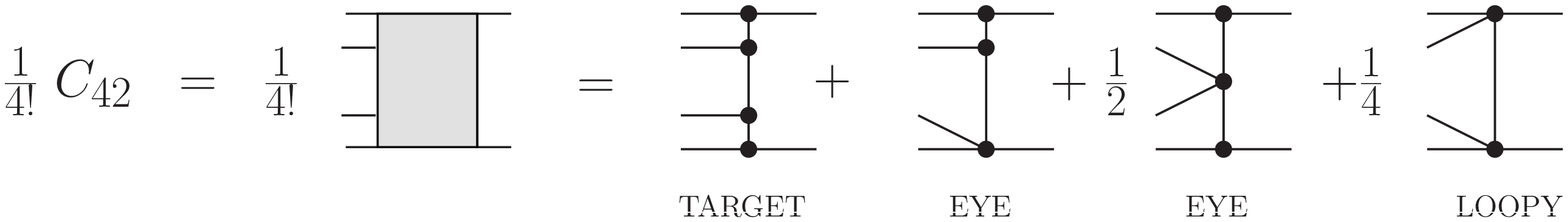}
\end{center}
\caption{\label{C42eqn}The shaded box is the vertex $C_{42}$ which is part of the kernel of the $\Omega$ integral equation. The vertex $C_{24}$ which appears in the $\Theta$ integral equation is obtained by inverting the diagram left to right.}
\end{figure}
\par\begin{figure}[H]
\begin{center}
\includegraphics[width=8cm]{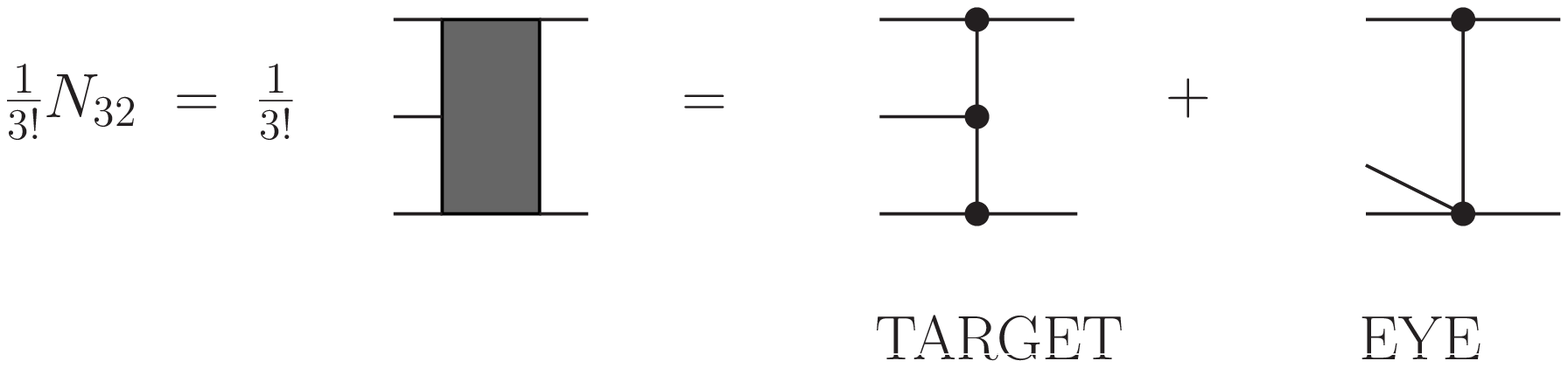}
\end{center}
\caption{\label{N32eqn}The shaded box is the vertex $N_{32}$ which is part of the kernel of the $\Psi$ integral equation. The vertex $N_{23}$ which appears in the $\Theta$ integral equation is obtained by inverting the diagram left to right.}
\end{figure}


\section{Integral Equations for Variational Vertex Functions}
\label{sectionUV}

The integral equations derived in the previous section for the vertex functions $\Omega$ and $\Psi$ depend on the vertices $U$ and $V$. These vertices satisfy the integral equations shown in Fig. \ref{Uint}, which are obtained from the equations of motion (\ref{eom1}). In this figure, we have combined diagrams that correspond to permutations of external legs.

\par\begin{figure}[H]
\begin{center}
\includegraphics[width=16cm]{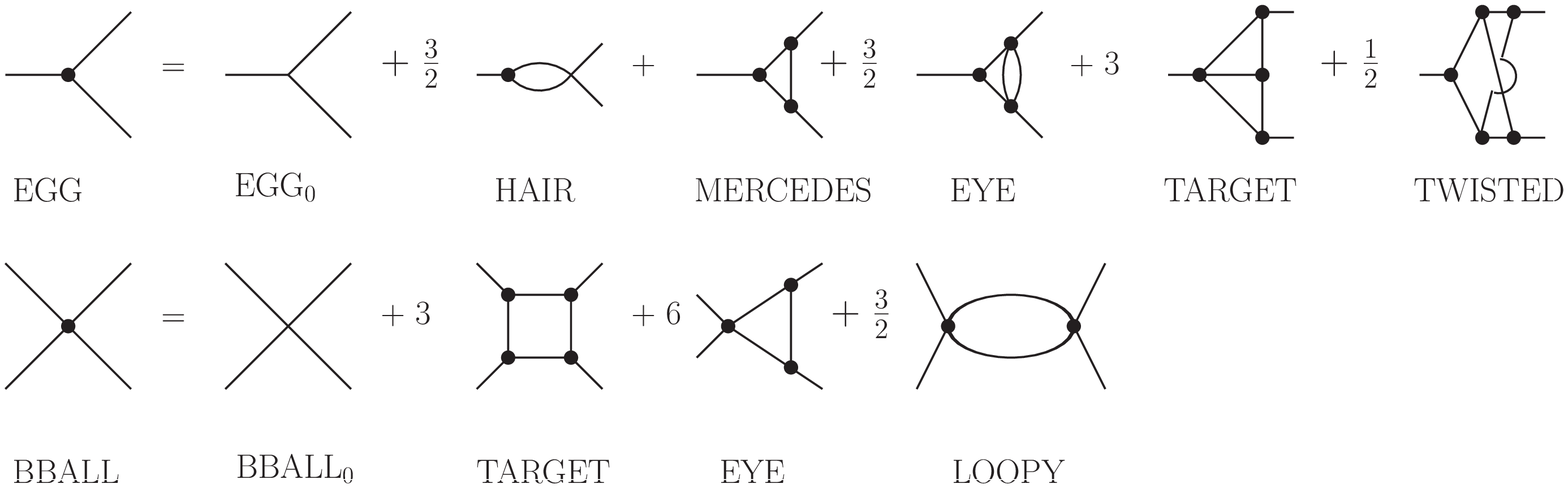}
\end{center}
\caption{\label{Uint}Integral equations for the vertices $U$ and $V$.}
\end{figure}

\section{Conclusion}
\label{Conc}

In this paper we have calculated the 4-loop 4PI effective action (Eq. (\ref{Gamma4PI}) and Fig. \ref{PHI}) for a scalar theory
with cubic and quartic interactions, with a non-vanishing field expectation value. We derive a set of coupled integral equations that give the corresponding re-summed 2-point vertex function. The resulting expression has the same form as the Schwinger Dyson equation (Eq. (\ref{Dext-3}) and Fig. \ref{PIextdiag}).  The Kubo formulae relate transport coefficients to 2-point correlators. A next-to-leading order contribution to the shear viscosity could be obtained by solving the set of equations derived in this paper.  We have checked that the equations derived in this paper are correct by expanding the re-summed 2-point vertex function to 3-loop order and verifying that all contributions are present, with the correct symmetry factors \cite{MC-CP}. \\

\noindent {\bf Acknowledgements}: 
The authors thank Julien Serreau for many useful comments and suggestions. \\

\appendix

\section{}
\label{appendixA}
In this appendix we present some details of our derivation of the 4PI effective action for a scalar theory
with cubic and quartic interactions, with a non-vanishing field expectation value. We 
follow the method of Ref. \cite{berges1}. We use a symbolic notation that suppresses indices, and therefore combines terms that correspond to permutations of indices. For example, a sum of three terms that corresponds to the symmetric combination of the product of a propagator and a field expectation value is written:
\bea
D_{ij}\phi_k+D_{jk}\phi_i+D_{ki}\phi_j~\rightarrow~3D\,\phi\,.
\eea
We start with the generating functional for 
Green's functions in the presence of quadratic, 
cubic and quartic source terms which is given by:
\begin{eqnarray}
\label{genFcn}
  Z[J, R, R_3, R_4]&=&{\rm Exp}\big[i W[J, R, R_3, R_4]\big]\nonumber \\ 
  &=&
  \int D\hat\phi \;{\rm Exp}\Big[i \Big(S_{cl}[\hat\phi]+J \hat\phi+\frac{1}{\;2!}R \hat\phi^2 +\frac{1}{\;3!}R_3 \hat\phi^3+\frac{1}{\;4!}R_4 \hat\phi^4\Big) \Big].
\end{eqnarray}
Derivatives of the generating functional $W$
with respect to sources produce connected 2-, 3- and 4-point functions 
which, in turn, can be expressed in terms of the proper 3- and 4-point vertices.

In this appendix, we introduce some new notation for the proper vertices. First, we remind the reader that the definitions in Eqs. (\ref{scl}) and (\ref{free}) were chosen to make figures look as simple as possible: 
lines, and intersections of lines, correspond directly to these definitions for the propagator and vertices, with no additional factors of plus or minus $i$. In this appendix we use a different definition so that the Legendre transformation has a simpler form. The new definition is obtained from the old one by multiplying by $i$, and denoted with a boldface character:
\bea
\label{bold}
\{{\bf U}_{oo},{\bf U}_0,{\bf V}_0,{\bf U},{\bf V}\} = \{i U_{oo},iU_0,iV_0,iU,iV\}\,.
\eea
This notation is illustrated for the bare 3-point vertex in Fig. \ref{U0illustration}.
\par\begin{figure}[H]
\begin{center}
\includegraphics[width=11cm]{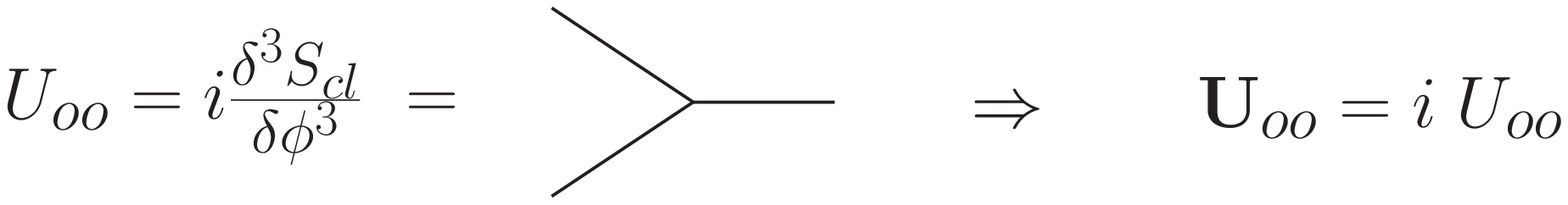}
\end{center}
\caption{\label{U0illustration}Illustration of the notation defined in (\ref{bold}) which will be used for all vertices in Appendix A. }
\end{figure}
\noindent We will use some additional vertices that are defined in terms of the boldfaced ones in Eq. (\ref{bold}) (see Eqs (\ref{modif_vert})). These additional vertices are used only to organize the calculation in this appendix and never appear in the body of the article, so we do not need to boldface them. 

We take functional derivatives of the generating functional $W$ defined in (\ref{genFcn}) with respect to the sources, and express the result in terms of the proper vertices defined in (\ref{bold}). We obtain:
\begin{eqnarray}
  \label{one}
  &&\frac{\delta W}{\delta J}=\phi\,, \nonumber\\
  &&\frac{\delta W}{\delta R}=\frac{1}{2}(D+\phi^2)\,, \nonumber\\
  &&\frac{\delta W}{\delta R_3}=\frac{1}{\;3!}(-iD^3 {\bf U}+3D\phi+\phi^3)\,,\nonumber\\
  \label{five}
  &&\frac{\delta W}{\delta R_4}=
  \frac{1}{\;4!}(-i D^4 {\bf V} - 3  D^5 {\bf U}^2-i D^3 {\bf U} \phi+3D^2+6D \phi^2+\phi^4)\,. 
\end{eqnarray}

The 4PI effective action is the Legendre transform of the generating function $W[J, R, R_3, R_4]$:
\begin{eqnarray}
\label{Legendre}
\Gamma[\phi, D, {\bf U}, {\bf V}]&=&W[J, R, R_3, R_4]-\frac{\delta W}{\delta J}J-
  \frac{\delta W}{\delta R}R-\frac{\delta W}{\delta R_3}R_3-\frac{\delta W}{\delta R_4}R_4\,.
\end{eqnarray}
It is not necessary to perform the Legendre transforms in $(\ref{Legendre})$
all at once: it is possible to implement them successively, expressing
higher-order effective actions in terms of lower order ones. The
procedure is as follows.
We start with the 2PI effective action for a scalar theory with cubic
and quartic interactions, with a non-vanishing field expectation value:
\begin{eqnarray}
\label{2PI}
  \Gamma[\phi, D]&=&S[\phi]+\frac{i}{2} {\rm Tr} {\rm Ln} D^{-1}
  +
\frac{i}{2} {\rm Tr}\left[(D^0_{12}(\phi))^{-1}\left(D_{21}-D^0_{21}(\phi)\right)\right]
+\Gamma_2[\phi, D]\,.
\end{eqnarray}
%
%
%
Since we are working to 4-loop order, $\Gamma_2[\phi, D]$ contains  all 2PI diagrams with 2, 3 or 4-loops. 
We note that the  $\phi$-dependence of $\Gamma_2[\phi, D]$ can be written 
as a function of the effective classical vertex defined in Section \ref{section4PI}. In the notation used in this appendix we have: ${\bf U}_0:={\bf U}_{oo}+\phi {\bf V}_0$.
Diagrams that contribute to $\Gamma_2[\phi, D]$ have vertices given by ${\bf U}_0$ and ${\bf V}_0$, 
and lines are the self consistent propagator $D$.

The strategy of the calculation is to define a 2PI effective action with a modified interaction:
\bea
\label{2PI-transform}
\Gamma_{\bar{U}\bar{V}}[\phi, D]&=&W[J, R, R_3, R_4]-\frac{\delta W}{\delta J}J-
  \frac{\delta W}{\delta R}R\,.
\eea
The subscripts on $\Gamma_{\bar{U}\bar{V}}[\phi, D]$ indicate that it depends on the sources $R_3$ and $R_4$ only through the modified interaction vertices which are defined as: 
\begin{eqnarray}
\label{modif_vert}
  && \bar{U}:=  {\bf U}_{oo}-R_3\,,~~~~\bar{V}:={\bf V}_{0}-R_4\,.
\end{eqnarray}
The structure of (\ref{2PI-transform}) is a consequence of the fact that the cubic and quartic source terms 
in Eq. (\ref{genFcn}) can be combined
with the 3- and 4-point interaction terms in Eq. (\ref{scl}) by making the definitions in Eq. (\ref{modif_vert}). The effective action defined in (\ref{2PI-transform}) has exactly the same form as the effective action in (\ref{2PI}), with the vertices ${\bf U}_{oo}$ and ${\bf V}_{0}$ replaced by $\bar U$ and $\bar V$, respectively.

The next step is to express the 4PI effective action in  ($\ref{Legendre}$) completely
in terms of the vertices ${\bf U}_{oo}$, ${\bf V}_0$, $\bar{U}$, $\bar{V}$ and the 2PI
effective action $\Gamma_{\bar{U}\bar{V}}[\phi, D]$. This is accomplished using the relations:
\begin{eqnarray}
  \label{exact}
  \frac{\delta \Gamma_{\bar U\bar{V}}}{\delta R_3}=\frac{\delta W}{\delta R_3}\,, \ \ 
  \frac{\delta \Gamma_{\bar U\bar{V}}}{\delta R_4}=\frac{\delta W}{\delta R_4}\,,
\end{eqnarray}
which are obtained from (\ref{2PI-transform}).
Using the definitions (\ref{modif_vert}) we obtain:
\begin{eqnarray}
\label{4PI}
\Gamma[\phi, D, {\bf U}, {\bf V}]&=&\Gamma_{\bar{U}\bar{V}}[\phi, D]-
  \frac{\delta W}{\delta R_3}R_3-\frac{\delta W}{\delta R_4}R_4 \nonumber\\ 
  &=&\Gamma_{\bar U\bar{V}}[\phi, D]-
  \frac{\delta \Gamma_{\bar{U}\bar{V}}[\phi, D]}{\delta \bar{U}}(\bar{U}-{\bf U}_{oo}) -\frac{\delta \Gamma_{\bar{U}\bar{V}}[\phi, D]}{\delta \bar{V}}(\bar{V}-{\bf V}_{0}).    
\end{eqnarray}

The last step is to to express the modified vertices $\bar{U}$ and $\bar{V}$
in terms of ${\bf U}$ and ${\bf V}$. This is done by comparing two expressions that relate these vertices.

On one hand, using (\ref{one}) and (\ref{exact})
one obtains the expressions for the derivatives of the modified
2PI effective action $\Gamma_{\bar U\bar{V}}[\phi, D]$:
\begin{eqnarray}
\label{exact_1}
  \frac{\delta \Gamma_{\bar{U}\bar{V}}[\phi, D]}{\delta \bar{U}}&=&
-  \frac{1}{\;3!}(-iD^3 {\bf U}+3D\phi+\phi^3)\,, \nonumber\\
  \label{exact_2}
  \frac{\delta \Gamma_{\bar{U}\bar{V}}[\phi, D]}{\delta \bar{V}}&=&
-  \frac{1}{\;4!}(-i D^4 {\bf V} - 3  D^5 {\bf U}^2-i D^3 {\bf U} \phi+3D^2+6D \phi^2+\phi^4)\,. 
\end{eqnarray}
On the other hand, one can explicitly take derivatives of the 4-loop 2PI effective action with bare vertices replaced by the modified interaction vertices (Eq. (\ref{modif_vert})). The non-4PI contributions to the interacting part are drawn in Fig. \ref{PHIbar} (see Ref. \cite{Pelster}).
\par\begin{figure}[H]
\begin{center}
\includegraphics[width=12cm]{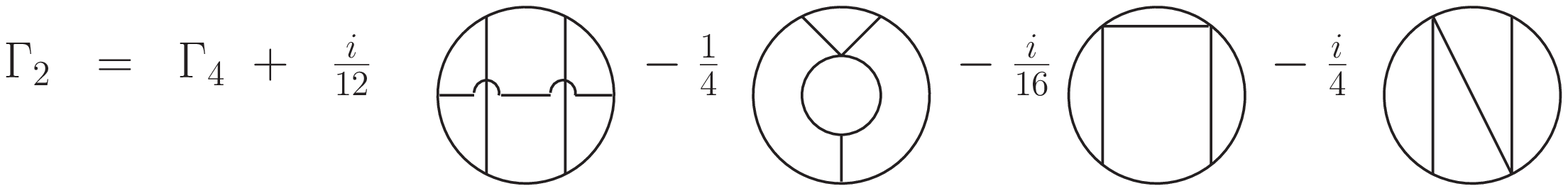}
\end{center}
\caption{\label{PHIbar}The interacting part of the modified 2PI effective action $\Gamma_{\bar U\bar{V}}[\phi, D]$ to 4-loop order. The 3-point vertices are $\overline{U}:=\bar U+\phi\bar V$ and the 4-point vertices are $\bar V$. The 4PI terms are denoted $\Gamma_4$. They are shown in Fig. \ref{PHI} using the vertices defined in (\ref{free}) and are not redrawn here. }
\end{figure}
\noindent Taking derivatives of the 4-loop expansion of $\Gamma_{\bar{U}\bar{V}}$ we obtain:
\bea
\label{exact_34}
\frac{\delta \Gamma_{\bar U\bar{V}}[\phi, D]}{\delta \bar U}&& = -\frac{\phi^3}{6}-\frac{1}{2}D\phi+\frac{i}{6}D^3\,\overline{ U}+\frac{1}{4}D^5\bar V\overline{U}\nonumber\\
&& -\frac{5i}{8} D^7 \bar V^2 \overline{ U}-\frac{i}{6}D^6\overline{U}^3 -\frac{3}{2}D^8\bar V \overline{U}^3+\frac{7i}{12} D^9 \overline{U}^5\,,\nonumber \\
 \frac{\delta \Gamma_{\bar U\bar{V}}[\phi, D]}{\delta \bar V} && =-\frac{1}{24}\phi^4 -\frac{1}{4}D\phi^2 -\frac{1}{8}D^2+\frac{i}{24}D^4\bar V+\frac{1}{16}D^6\bar V^2  + \frac{i}{6}D^3\phi \overline{U}\nonumber\\
 && + \frac{1}{4} D^5 \phi\bar V \overline{U}
-\frac{5i}{8}D^7\phi\bar V^2
 \overline{U}+\frac{1}{8}D^5\overline{U}^2-\frac{5i}{8}D^7\bar V\overline U^2
 -\frac{i}{6}D^6\phi \overline{U}^3\nonumber\\
 && -\frac{3}{2}D^8 \bar V\phi \overline{ U}^3-\frac{3}{8}D^8 \overline{U}^4+\frac{7i}{12} D^9\phi \overline{U}^5 \,.
\eea
Equating (\ref{exact_1}) and (\ref{exact_34}) and using an iterative procedure to rearrange, we get the results in Eq. (\ref{iterated}), where the dots indicate that we have dropped contributions that are of higher loop order. 
\bea
\label{iterated}
&&\bar V={\bf V} + \frac{3i}{2}D^2 {\bf V}^2+6D^3{\bf V}{\bf U}^2-3iD^4 {\bf U}^4 +\cdots\nonumber\\
&&\overline{U}={\bf U}+\frac{3i}{2}D^2 {\bf U} {\bf V}+D^3 {\bf U}^3-\frac{3}{4}D^4 {\bf U} {\bf V}^2+6i D^5 {\bf U}^3 {\bf V}+4 D^6 {\bf U}^5 + \cdots
\eea
Substituting (\ref{exact_1}) and (\ref{iterated}) into (\ref{4PI}) and using (\ref{2PI})  we obtain from a straightforward calculation:
\begin{eqnarray}
\Gamma[\phi, D, {\bf U}, {\bf V}]&& = S[\phi]+\frac{i}{2} {\rm Tr} {\rm Ln} D^{-1}+
\frac{i}{2} {\rm Tr}\left[(D^0_{12}(\phi))^{-1}\left(D_{21}-D^0_{21}(\phi)\right)\right] \nonumber\\
&& + \Gamma^0[\phi, D, {\bf U}, {\bf V}]+\Gamma^{int}[D, {\bf U}, {\bf V}]\,,\nonumber\\
\Gamma^0[\phi, D,{\bf U}, {\bf V}] && = -\frac{1}{8}  D^2 {\bf V}_{0} +
  \frac{i}{6}  D^3 {\bf U}_{0}{\bf U} + \frac{i}{24} D^4  {\bf V} {\bf V}_{0}+\frac{1}{8}  D^5 {\bf V}_{0}{\bf U}^2\,,\nonumber\\
 \Gamma^{int}[ D, {\bf U}, {\bf V}] 
&&= -\frac{i}{12} D^3 {\bf U}^2-\frac{i}{48} D^4 {\bf V}^2
  -\frac{i}{24} D^6  {\bf U}^4 + \frac{1}{48}  D^6 {\bf U}^3-\frac{i}{8} D^7 {\bf V}^2 {\bf U}^2\nonumber\\
&&  -\frac{1}{8} D^8  {\bf U}^4{\bf V}+\frac{i}{72} D^9 {\bf U}^6. 
\end{eqnarray}
Converting the notation for the vertices using Eq. (\ref{bold}), these results for $\Gamma^0[\phi, D, V_3, V_4]+\Gamma^{int}[D, V_3, V_4]$ are shown in Fig. \ref{PHI}. Thus we find that the Legendre transform has removed the non-4PI terms that were present in the modified 2PI effective action (\ref{2PI-transform}).

\end{document}